\documentclass[aps,showpacs,twocolumn,twoside,pre,superscriptaddress,amsmath,amssymb,amsfonts,10pt]{revtex4-2}
\usepackage[utf8]{inputenc}
\usepackage{epsfig}
\usepackage{graphicx}
\usepackage{array}
\usepackage{xcolor}
\usepackage{hyperref}

\hypersetup{
    colorlinks=true,
    linkcolor=blue,
    filecolor=magenta,      
    urlcolor=cyan,
    pdftitle={Overleaf Example},
    pdfpagemode=FullScreen,
    }

\def\ket#1{\left|{#1}\right\rangle}
\def\freg{f_{\mathrm{reg}}}
\def\lyapunov{\lambda_{\mathrm{cl}}}
\def\lyapunove{\overline{\lambda}_{\mathrm{cl}}}
\def\Hcl{H_{\mathrm{cl}}}
\def\alg#1{\mathrm{#1}}
\def\op#1{\hat{#1}}

\def\vector#1{\mathbf{#1}}

\graphicspath{{figures/}}

\begin{document}
\title{Relative asymptotic oscillations of the out-of-time-ordered correlator as a quantum chaos indicator}

\author{Jakub Novotný}
\email{novotny.jakub@ipnp.mff.cuni.cz}
\affiliation{Institute of Particle and Nuclear Physics, Faculty of Mathematics and Physics, Charles University, V Hole\v{s}ovi\v{c}k\'ach 2, 18000 Prague, Czech Republic} 
\author{Pavel Stránský}
\email{stransky@ipnp.troja.mff.cuni.cz}
\affiliation{Institute of Particle and Nuclear Physics, Faculty of Mathematics and Physics, Charles University, V Hole\v{s}ovi\v{c}k\'ach 2, 18000 Prague, Czech Republic} 

\date{\today}

\begin{abstract}
A detailed numerical study reveals that the asymptotic values of the standard deviation-to-mean ratio of the out-of-time-ordered correlator in energy eigenstates can be successfully used as a measure of the quantum chaoticity of the system.
We employ a finite-size fully connected quantum system with two degrees of freedom, namely the algebraic $\alg{u}(3)$ model, and demonstrate a clear correspondence between the energy-smoothed relative oscillations of the correlators and the ratio of the chaotic part of the volume of phase space in the classical limit of the system.
We also show how the relative oscillations scale with the system size and conjecture that the scaling exponent can also serve as a chaos indicator.
\end{abstract}

\maketitle

\section{Introduction}
The correspondence between the theory of classical and quantum chaos has been extensively studied since the formulation of the Bohigas-Giannoni-Schmit conjecture~\cite{Bohigas1984}.
While classical chaos is rigorously constructed mathematically and routinely studied, most often by the rate of exponential divergence of neighboring trajectories, the study of quantum chaos is more intriguing.
Quantum chaoticity is usually defined indirectly by a comparison of suitable properties of a quantum system---most often correlations in energy spectra---with the chaoticity of its classical counterpart. In particular scenarios, the theory of classical-quantum correspondence is well-established~\cite{Gutzwiller1990,Haake2010}.

In this paper, we move from the static description of the quantum chaos based on spectral properties to the dynamical manifestations.
We will employ the Out-of-Time-Ordered Correlators (OTOCs)---four-point correlation functions of two quantum operators taken at different times.

The OTOCs have already gained popularity as an indicator of quantum chaos, especially their short-time evolution due to their connection with classical instability.
They were introduced long ago as a semiclassical tool to study superconductivity~\cite{Larkin1969} and later dusted off by showing their relevance in black-hole physics and chaos~\cite{Shenker2014,Maldacena2016}.
In the case of a quantum system with the classical limit, the short-time behavior of the OTOCs mimics the exponential spreading of neighboring classical trajectories up to the Ehrenfest time, leading to the notion of quantum Lyapunov exponent~\cite{Rozenbaum2017,GarciaMata2018,Rozenbaum2019,Fortes2019,Hummel2019,ChavezCarlos2019,Huang2019,Pappalardi2020,PilatowskyCameo2020,Craps2020,Sinha2021,Morita2022,GarciaMata2022} and quantum butterfly effect~\cite{Shenker2014,Roberts2016,Hosur2016,Dora2017,Khemani2018,Morita2022}.
In quantum systems with local interactions and local OTOC operators, the initial time evolution of the OTOCs describes the spreading (or scrambling) of quantum information~\cite{Swingle2016,Swingle2018,Borgonovi2019,Riddell2019,GonzalezAlonso2019,Yan2020,Mi2021}.
Based on the rate of initial growth of OTOCs, independently of the existence of the classical limit, quantum systems can be placed into a category of slow scramblers (maximally polynomial initial growth even in the presence of chaos)~\cite{Kukuljan2017,Lin2018,Marino2019,Fortes2019}, or fast scramblers (exponential scrambling)~\cite{Roberts2016,GarciaMata2018,ChavezCarlos2019,Fortes2019}.
The black holes have been conjectured as the fastest scramblers~\cite{Shenker2014}, employing the AdS/CFT duality~\cite{Sekino2008}.
OTOC analysis is being applied for studying many-body quantum systems, spin systems, and quantum circuits~\cite{Mi2021}. In recent years, this vast area for application gave rise to many proposals and measuring protocols~\cite{Swingle2016,Landsman2019,LewisSwan2019,Niknam2020,Sanchez2020,Rautenberg2020,Green2022,Blocher2022} followed by experimental measurements~\cite{Gaerttner2017,Mi2021,Braumueller2022}.
OTOCs have also been used to study many-body quantum scars~\cite{Sinha2021,Yuan2022} and quantum thermalization~\cite{Borgonovi2019,Kidd2021}, and as a probe of the excited-state quantum phase transitions~\cite{Wang2019}.

Whereas there is a vast amount of literature about the study of the short-time OTOC evolution, a possible connection of its long-time properties to chaos has been hinted at in a few recent works.
In the case of finite-size many-body quantum systems, OTOCs saturate in the long-time regime and exhibit oscillations around the mean value.
In Ref.~\cite{GarciaMata2018}, it was suggested for the first time that the appearance of saturation oscillations could be linked to quantum chaos. 
Later, a numerical study of a quantum Harper map demonstrated that the suppression of the OTOC oscillations in infinite temperature limit quantitatively corresponds to the ratio of the chaotic volume of the whole phase space~\cite{Fortes2019}. 
The same authors have shown~\cite{Fortes2020} that the suppression of relative oscillations of infinite-temperature OTOC can reveal the transition to chaos even in a small chain of four spins with OTOC studied on time scales attainable by experiment~\cite{Li2017}. 
The relative oscillations of OTOCs as a qualitative chaos indicator have also been mentioned in other studies~\cite{Anand2021,Goto2021}.
Some studies propose that even the asymptotic OTOC mean value can serve as a chaos indicator~\cite{Sinha2021}. 

The short-time exponential behavior of OTOCs is not a bulletproof sign of quantum chaoticity. 
Unstable potential stationary points can cause exponential growth of OTOCs even in integrable systems~\cite{Rozenbaum2020,PilatowskyCameo2020,Xu2020,Wang2021a,Kidd2021}, hence misleadingly indicating chaos. 
And vice versa, the already mentioned slow scramblers show polynomial growth of OTOC even in chaotic regimes~\cite{Fortes2019}. 
In this context, one can ask to what extent the relative oscillations overcome the drawbacks of the short-time OTOCs.
Another point can be whether they allow not only for indicating chaos but also for measuring it, and how well they compare to other known chaos measures.
Some clues are already present in the literature. 
Authors in Ref.~\cite{Kidd2021} noticed that the size of oscillations of OTOCs in the long-time regime in the simple Bose-Hubbard dimer could qualitatively distinguish nonchaotic, mixed, and fully chaotic regions, while the OTOCs show exponential growth even in the fully regular regime of the model. 
And in the already mentioned work~\cite{Fortes2020}, authors show that chaos suppresses relative oscillations of OTOCs even in the slowly scrambling spin chains. 

In this paper, we will be interested mainly in the long-time behavior of OTOCs, although we will also briefly discuss their short-time evolution and compare it with classical stability.
We will work with the eigenenergy expectation values of the OTOCs, which will allow us to analyze the energy dependence of relative oscillations and link them to other well-established energy-dependent measures of chaoticity. 
Specifically, we will show that the ratio of the asymptotic standard deviation to the asymptotic mean value (called wiggliness for brevity) compares to the volume of the regular part of the phase space in the classical limit. 

We will provide a detailed numerical study in a bosonic algebraic quantum model based on the $\alg{u}(3)$ algebra.
This model has been originally introduced to describe bending modes of linear polyatomic molecules~\cite{Iachello1996,SanchezCastellanos2009,EstevezFregoso2018,BermudezMontana2020}.
Later it has been successfully used to investigate purely theoretical concepts, such as quantum monodromy and quantum critical effects~\cite{PerezBernal2008,Larese2011,KhaloufRivera2022}.
Recently it has also been applied to study the properties of spinor Bose-Einstein condesates~\cite{Gerving2012,Rautenberg2020,Feldmann2021,Cabedo2021}, which are experimentally achievable, for example, as a condensate of cold rubidium atoms~\cite{Kunkel2018,Kunkel2019}.
Some aspects of the OTOCs and their relation to chaos have already been studied in this model~\cite{Rautenberg2020}.

We have chosen the $\alg{u}(3)$ model because its Hilbert space is finite. 
The model has just 2 degrees of freedom and a known classical limit.
On top of that, it is nonintegrable with enough parameters to tune the chaoticity.
Due to the algebraic origin, all observables can be easily constructed from the operators of the $\alg{u}(3)$ algebra.
And finally, one can vary the number of boson excitations in the system and study how the asymptotic OTOC behavior changes with the system size.

This paper is organized as follows.
In Sec.~\ref{sec:Theory} we recall some necessary concepts of the theory of classical chaos; we introduce the OTOCs and the wiggliness, which is the primary tool of the present work, and briefly discuss the correspondence between the classical and quantum chaos indicators. Sec.~\ref{sec:Model} introduces the $\alg{u}(3)$ model and its classical limit. The numerical results and their discussion is given in Sec.~\ref{sec:Results}. We finally conclude in Sec.~\ref{sec:Conclusions}.

\section{Theory}
\label{sec:Theory}
In this section, we introduce the classical Lyapunov exponent and the fraction of the regular part of the energy hypersurface in the phase space. Then we present energy-dependent microcanonical OTOCs.
Finally, we identify short-time and long-time evolution in the OTOC dynamics and specify to which classical notion they can be compared.


\subsection{Classical dynamics}
We consider a classical system with $f$ degrees of freedom described by time-independent Hamiltonian $H_{\mathrm{cl}}(\vector{x})$, which is a function of coordinates and conjugated momenta ${\vector{x} \equiv (x_1,\dots, x_{2f})=(p_{1},\dotsc,p_{f},q_{1},\dotsc,q_{f})}$ on a phase space. 
The time independence of the Hamiltonian implies energy conservation. 
Therefore, classical trajectories are constrained to $(2f-1)$-dimensional hypersurfaces $\Sigma_E$ determined by constant energy $\Hcl=E$.

One of the prominent signs of classical chaos is the exponential separation of asymptotically close trajectories characterized by the positivity of the Lyapunov exponent~\cite{Skokos2010,Lerose2020}
\begin{equation}
\label{eqn:LyapExp}    
    \lyapunov(\vector{x}) = \lim_{t \rightarrow \infty}\lim_{|\delta \mathbf{x}| \rightarrow 0}\frac{1}{t}\log\frac{|(\mathbf{x} + \delta \mathbf{x})(t) - \mathbf{x}(t)|}{|\delta \mathbf{x}|} ,
\end{equation}
where $\vector{x}(t)$ and $(\vector{x} + \delta \vector{x})(t)$ are trajectories with initial conditions $\vector{x_{0}}$ and $\vector{x}_0 + \delta \vector{x}$, respectively, $\delta \mathbf{x}$ is an infinitesimally small deviation vector in the neighbourhood of the point $\mathbf{x}_0$ on the given energy hypersurface, and $\left|\bullet\right|$ is a norm in the phase space.
In the case of regular (stable) trajectories $\lyapunov=0$.

Apart from the Lyapunov exponent $\lyapunov(\vector{x})$ characterizing the stability of a single trajectory, we will employ the average Lyapunov exponent on the energy hypersurface $\Sigma_{E}$ calculated as
\begin{equation}
    \label{eq:LyapE}
    \overline{\lambda}_{\mathrm{cl}}(E)=\frac{\int\delta(E-H(\vector{x}))\lambda_{\mathrm{cl}}(\vector{x})d^{2f}\vector{x}}{\Gamma(E)},
\end{equation}
where $\Gamma(E)=\int\delta(E-H(\vector{x}))d^{2f}\vector{x}$ is the entire volume of $\Sigma_{E}$.

The Lyapunov exponent explains the local properties of the dynamics. 
On the other hand, the overall chaoticity of the system with energy $E$ is reflected in the fraction of regularity
\begin{equation}
    \label{eqn:f_regDefinition}
    \freg(E) = \frac{\Gamma_{\mathrm{reg}}(E)}{\Gamma(E)}\in[0,1],
\end{equation}
where $\Gamma_{\mathrm{reg}}(E)$ is the volume of all the regular regions in $\Sigma_E$, which, in general, are well-separated from the regions of chaotic dynamics.
The limit values correspond to fully chaotic ($\freg=0$) and fully regular ($\freg=1$) dynamics at the given energy $E$.

\subsection{Quantum dynamics}
Quantum chaos is often studied from static properties of the spectral correlations~\cite{Haake2010}.
Here we focus on the dynamical manifestations, namely on the properties of the OTOC. 
In a quantum system described by Hamiltonian $\hat{H}$, 
the out-of-time-ordered correlator is introduced as the expectation value in energy eigenstates $\ket{E_{n}}$
\begin{equation}
    \label{eqn:MicroOTOCs}
    C_n(t) = \left\langle E_{n} \middle| [\hat{V}(t),\hat{W}(0)]^{\dagger}[\hat{V}(t),\hat{W}(0)] \middle|E_{n}\right\rangle,
\end{equation}
where $E_{n}$ is the $n$-th eigenenergy, $\hat{H}\ket{E_{n}} = E_{n}\ket{E_{n}}$, and $\hat{V},\hat{W}$ are for the moment arbitrary quantum operators in the Heisenberg picture,
\begin{equation}
    \label{eqn:Heisenberg}
    \hat{V}(t) = e^{\frac{i}{\hbar}\hat{H}t} \hat{V} e^{-\frac{i}{\hbar}\hat{H}t};
\end{equation}
$[\hat{V}(t),\hat{W}(0)]$ is their commutator.
If both operators are Hermitian, then 
\begin{equation}
    \label{eqn:MicroOTOCHerm}
    C_n(t) = -\left\langle E_{n}\middle|\left[\hat{V}(t),\hat{W}(0)\right]^2\middle| E_{n} \right\rangle.
\end{equation}
The OTOC properties undoubtedly depend on the choice of the operators. 
These are usually relevant physical observables of the system, such as positions, momenta operators, or local spin operators in models with finite-range interaction.
The variability of the OTOC dynamics for several pairs of operators will be demonstrated later in Sec.~\ref{sec:Results}.

Note that the OTOCs are often considered as thermal averages in the canonical ensemble at inverse temperature $\beta$~\cite{Maldacena2016,Roberts2016},
 sometimes even at infinite temperature $\beta=0$ only~\cite{GarciaMata2018,Fortes2019}.
For the sake of distinguishing pure state and thermal state averaging, the OTOCs of type~\eqref{eqn:MicroOTOCs} are 
called microcanonical OTOCs in the literature~\cite{Hashimoto2017,PilatowskyCameo2020}. 

\subsection{Classical-quantum correspondence}\label{sec:CQC}
The time dependence of a general OTOC can be divided into short-time and long-time regimes, roughly limited by the Ehrenfest (or scrambling) time $t_{E}\propto\lyapunov^{-1}\ln{N}$~\cite{Pappalardi2018,Rammensee2018,Fortes2019,GarciaMata2022,Richter2022}, where $N$ is the size parameter of the system.
In more detailed analyses, there has also been described a universal power-law growth at very short times $t_{d}<t_{E}$~\cite{Pappalardi2020} ($t_{d}$ is called the dephasing time) for OTOC operators commuting at $t=0$, and yet another time scale given by the diffusion time $t_{D}\gg t_{E}$, which is the time of complete saturation of the quantum dynamics~\cite{Benet2022}.

Whereas the OTOCs in eigenstates corresponding to regular dynamics typically show strong oscillations at all times, OTOCs for chaotic eigenstates are characterized by initial exponential growth for $t<t_{E}$ and long-time saturation regime with small aperiodic fluctuations around the mean value for $t>t_{D}$.
In the chaotic systems with a classical limit, the exponent $\lambda$ of the initial exponential growth 
\begin{equation}
    \label{eq:InitialGrowth}
    C(t) \propto e^{2\lambda t}
\end{equation}
is related to the classical Lyapunov exponent $\lyapunov$, hence often dubbed quantum Lyapunov exponent. Especially, for operators $\hat{x}_i$ that correspond to the canonical coordinates and momenta $x_i$ on the classical phase space the exponent $\lambda$ coincides with the Lyapunov exponent~\cite{Larkin1969,Hashimoto2017,ChavezCarlos2019},
\begin{equation}
\label{eq:QuantumLyap}
    [\hat{x}_i(t),\hat{x}_j(0)]^2 \leftrightarrow \{x_i(t),x_j(0)\}^2 \sim e^{2\lyapunov t},
\end{equation}
where $\{\bullet,\bullet\}$ are the classical Poisson brackets; the indices $i,j=1,\dotsc,2f$ can be taken arbitrarily.

In general, however, the equality $\lambda=\lyapunov$ is not valid for all choices of OTOC operators.
It can be shown that the Poisson brackets $\{A(t),A(0)\}$ of a general function $A$ on the phase space have a Lyapunov term that mimics the exponential separation of infinitesimally close chaotic trajectories and which is projected onto the gradient of~$A$. Therefore, this gradient and other terms containing the second derivative of $A$ may modify the exact value of $\lambda$, and the corresponding OTOC will not grow exactly with the classical Lyapunov exponent.

The long-time behavior of OTOCs for a given pair of operators $\hat{V}$, $\hat{W}$ is captured by the mean value and variance
\begin{align}
\label{eq:OTOCmean}
\overline{C}_n&=\lim _{T \rightarrow \infty} \frac{1}{T} \int_{0}^T C_n(t) dt,\\
\label{eq:OTOCvar}
\sigma^2_n&=\lim _{T \rightarrow \infty} \int_{0}^T C_n^{2}(t) dt-\overline{C}_n^2.
\end{align}
It will be shown later that the crucial role in quantifying chaoticity is played not by these quantities themselves but by their ratio (the coefficient of variation)
\begin{equation}\label{eq:OTOCFano}
\nu_n=\frac{\sigma_n}{\overline{C}_n},
\end{equation}
reflecting the relative oscillations of the OTOC.
In the following text, we will, for the sake of brevity, call this ratio~\emph{wiggliness}, and we will always specify for which operators the OTOCs and the wiggliness are computed.

\section{Model}\label{sec:Model}
The model Hamiltonian belongs to a class of boson-interacting systems. 
It is constructed from nine generators of the spectrum-generating $\alg{u}(3)$ Lie algebra~\cite{Iachello1996}.
The generators can be represented by the bilinear products of creation and annihilation operators $\hat{b}_{i}^{\dagger}\hat{b}_{j}, i,j=0,1,2$, where $\hat{b}_{0}\equiv\hat{\sigma}$ is the scalar boson operator and $\hat{b}_{1,2}\equiv\hat{\tau}_{1,2}$ form the pair of circular boson operators $\hat{\tau}_{\pm}=\left(\hat{\tau}_{1}\pm i\hat{\tau}_{2}\right)/\sqrt{2}$, all of them satisfying the boson commutation relations
\begin{align}
    \left[\hat{b}_{i},\hat{b}_{j}^{\dagger}\right]&=\delta_{ij},&
    \left[\hat{b}_{i},\hat{b}_{j}\right]&=\left[\hat{b}_{i}^{\dagger},\hat{b}_{j}^{\dagger}\right]=0.
\end{align} 
The Hilbert space is spanned over all single-particle states,
\begin{equation}
    \ket{n_\sigma,n_{+},n_{-}}\equiv\frac{1}{\mathcal{N}}\left(\hat{\sigma}^{\dagger}\right)^{n_{\sigma}}\left(\hat{\tau}_{+}^{\dagger}\right)^{n_{+}}\left(\hat{\tau}_{-}^{\dagger}\right)^{n_{-}}\ket{0},
\end{equation}
where $n_{\sigma},n_{\pm}=0,1,2,\dotsc$ are the numbers of the corresponding boson excitations, $\ket{0}$ is the vacuum state and $\mathcal{N}$ a normalizing factor.
Note that any operator constructed from the generators conserves the total number of boson excitations $\hat{N}=\hat{\sigma}^{\dagger}\hat{\sigma}+\hat{\tau}_{+}^{\dagger}\hat{\tau}_{+}+\hat{\tau}_{-}^{\dagger}\hat{\tau}_{-}$.
The number $N=n_{\sigma}+n_{\tau}$, where $n_{\tau}=n_{+}+n_{-}$, specifies the fully-symmetric irreducible representation of $\alg{u}(3)$ and dictates the dimensionality of the corresponding (finite) Hilbert space $\mathcal{H}$,
\begin{equation}
    \label{eq:dimension}
    \mathrm{dim}\,\mathcal{H} = \frac{1}{2}(N+1)(N+2).
\end{equation}
Therefore, $N$ serves as a tunable size parameter of the system.

It is convenient to construct another set of generators as linear combinations of $\hat{b}_{i}^{\dagger}\hat{b}_{j}$,
\begin{align} 
    \hat{n}_{\tau}&=\hat{\tau}_{+}^{\dagger}\hat{\tau}_{+}+\hat{\tau}_{-}^{\dagger}\hat{\tau}_{-},&
    \hat{n}_{s}&=\hat{\sigma}^{\dagger}\hat{\sigma},\nonumber\\
    \hat{D}_{\pm}&=\pm\sqrt{2}\left(\hat{\tau}_{\pm}^{\dagger}\hat{\sigma}-\hat{\sigma}^{\dagger}\hat{\tau}_{\mp}\right),&
    \hat{Q}_{\pm}&=\sqrt{2}\hat{\tau}_{\pm}^{\dagger}\hat{\tau}_{\mp},\\
    \hat{R}_{\pm}&=\sqrt{2}\left(\hat{\tau}_{\pm}^{\dagger}\hat{\sigma}+\hat{\sigma}^{\dagger}\hat{\tau}_{\mp}\right),&
    \hat{l}&=\hat{\tau}_{+}^{\dagger}\hat{\tau}_{+}-\hat{\tau}_{-}^{\dagger}\hat{\tau}_{-},\nonumber
\end{align}
where the following subsets are generators of subalgebras
\begin{subequations}
    \begin{align}
        \alg{o}(3)&=\text{span}\left\{\hat{D}_{\pm},\hat{l}\right\},\\
        \overline{\alg{o}}(3)&=\text{span}\left\{\hat{R}_{\pm},\hat{l}\right\},\\
        \alg{u}(2)&=\text{span}\left\{\hat{n}_{\tau},\hat{l},\hat{Q}_{\pm}\right\},\\
        \alg{o}(2)&=\text{span}\left\{\hat{l}\right\}
    \end{align}        
\end{subequations}
that admit three subalgebra chains,
\begin{subequations}
    \begin{align}
        \mathrm{I}:&\alg{u}(3)\supset \alg{u}(2)\supset \alg{o}(2),\\
        \mathrm{II}:&\alg{u}(3)\supset \alg{o}(3)\supset \alg{o}(2),\\
        \overline{\mathrm{II}}:&\alg{u}(3)\supset\overline{\alg{o}}(3)\supset \alg{o}(2).
    \end{align}        
\end{subequations}
The $\alg{o}(2)$ is the symmetry algebra reflecting the invariance of the system with respect to the rotation around the $z$-axis.
The last two chains are equivalent and interchangeable, connected via a unitary transformation. 
Depending on the choice of algebra $\alg{o}(3)$ or $\overline{\alg{o}}(3)$, it is talked about a momentum or a coordinate realization of the same Hamiltonian~\cite{EstevezFregoso2018}.

A common way to construct the $\alg{o}(3)$ Hamiltonian is via the Casimir operators of all the possible subalgebras.
Here we consider only the linear Casimir $\hat{\mathcal{C}}_{1}[\alg{u(2)}]$ of the $\alg{u}(2)$ algebra and the quadratic Casimir $\hat{\mathcal{C}}_{2}[\alg{o(3)}]$ of the $\alg{o}(3)$ algebra, and build the Hamiltonian in the following form
\begin{equation}
    \label{eq:VibronHam}
    \hat{H}_{0}=\frac{1-\xi}{N}\underbrace{\hat{\mathcal{C}}_{1}[\alg{u}(2)]}_{\hat{n}_{\tau}}-\frac{\xi}{N(N-1)}\underbrace{\hat{\mathcal{C}}_{2}[\alg{o}(3)]}_{\hat{D}^{2}},
\end{equation}
where
\begin{equation}
    \hat{D}^{2}\equiv\frac{1}{2}\left(\hat{D}_{+}\hat{D}_{-}+\hat{D}_{-}\hat{D}_{+}\right)+\hat{l}^{2}
\end{equation}
and $\xi\in[0,1]$ is an adjustable parameter.
The factor $N$ is added so that the Hamiltonian scales correctly in the infinite-size limit $N\rightarrow\infty$.

The system given by $\hat{H}_{0}$ has $f=2$ degrees of freedom and is integrable with Hamiltonian and angular momentum operator $\hat{l}^{2}$ being the two independent integrals of motion.
This Hamiltonian is frequently used to model a quantum phase transition between the $\alg{o}(2)$ (usually called symmetric) phase and the $\alg{o}(3)$ (called deformed or displaced) phase, occurring at $\xi=1/5$ in the infinite-size limit~\cite{PerezBernal2008}.

The integrability of $\hat{H}_{0}$ will be broken by violating the $\alg{o}(2)$ symmetry,
\begin{equation}
    \hat{H}=\hat{H}_{0}-\frac{\epsilon}{N}\hat{D}_{x},
    \label{eq:hamiltonian}
\end{equation}
where $\hat{D}_{x}=(\hat{D}_{+}+\hat{D}_{-})/2$ is the dipole operator [$\hat{D}_{y}$ would be defined as $\hat{D}_{y}=(\hat{D}_{+}+\hat{D}_{-})/2i$].
Note that $\hat{H}$ still has a discrete symmetry (parity), so its spectrum forms two invariant sets of states in the Hilbert space. A suitable basis for distinguishing these two invariant subspaces is given by vectors 
\begin{equation}
    \label{eq:SecondBasis}
        |N,n_{\tau},l_{\pm}\rangle \equiv \frac{1}{\sqrt{2}} \left( |N,n_{\tau},l\rangle \pm |N,n_{\tau},-l\rangle\right),
\end{equation}
where $l$ is taken as non-negative. Two subspaces $\mathcal{H}^{(1,2)}$ invariant under the action of the Hamiltonian~\eqref{eq:hamiltonian} have the form
\begin{subequations}
    \begin{align}
        \mathcal{H}^{(1)} &= \text{span}\{|N,n_{\tau},l_{+}\rangle_{l \text{-even}},|N,n_{\tau},l_{-}\rangle_{l \text{-odd}}\},\\
        \mathcal{H}^{(2)} &= \text{span}\{|N,n_{\tau},l_{+}\rangle_{l \text{-odd}},|N,n_{\tau},l_{-}\rangle_{l \text{-even}}\},
    \end{align}        
\end{subequations}
where $n_{\tau} = \{0,1,\dots N\}$ and $l = \{0,1,\dots,n_{\tau}\}$.
The full Hilbert space is $\mathcal{H}=\mathcal{H}^{(1)}\oplus\mathcal{H}^{(2)}$.

The perturbed model is integrable for particular values of $\xi = 0 \text{ and } 1$, which correspond to excluding one of the Casimir operators in \eqref{eq:VibronHam}.
Instead of $\op{l}$, other operators play the role of the second integral of motion. 
For $\xi = 0$, the Hamiltonian commutes with the operator $\op{n} + \op{Q}$ where $\op{Q} =\left( \op{Q}_+ + \op{Q}_-\right)/\sqrt{2}$, and for $\xi = 1$, the Hamiltonian
 is formed only by operators from algebra $\alg{o}(3)$ and commutes with both $\op{D}^2$ and $\op{D}_x$.

The classical limit of the Hamiltonian can be obtained by transforming the circular boson operators into the coordinate-momentum form
\begin{equation}
    \hat{b}_{j}=\sqrt{\frac{N}{2}}\left(\hat{q}_{j}+i\hat{p}_{j}\right),\quad j=1,2,
\end{equation}
using the Holstein-Primakoff transformation~\cite{Holstein1940,Macek2019} to eliminate the degree of freedom connected with the $\sigma$ boson due to the conservation of $N$, and performing the limit $N\rightarrow\infty$, after which the operators with commutation relations 
\begin{equation}
    \label{eq:Comut}
    \left[\hat{q}_{j},\hat{p}_{k}\right]=\frac{i}{N}\delta_{jk}
\end{equation}
serve as canonically conjugated coordinates and momenta. 
The corresponding classical Hamiltonian function reads as
\begin{align}
    H_{cl}\equiv\lim_{N\rightarrow\infty}&\hat{H}=\left(1-\xi\right)\frac{s^{2}}{2}\nonumber\\
    &-\xi\left[\left(p_{1}^{2}+p_{2}^{2}\right)\left(2-s^{2}\right)+\left(p_{1}q_{2}-q_{1}p_{2}\right)^{2}\right]\nonumber\\
    &-\epsilon p_{2}\sqrt{2-s^{2}}.
    \label{eq:hcl}
\end{align}
The four-dimensional phase space $(p_{1},p_{2},q_{1},q_{2})$ is compact, bounded by the condition $s^{2}\leq2$, where $s^{2}=p_{1}^{2}+p_{2}^{2}+q_{1}^{2}+q_{2}^{2}$.

Note that the commutation relations~\eqref{eq:Comut} imply that the size parameter $N$ also serves as an effective Planck constant $\hbar_{\mathrm{eff}}=1/N$. Therefore, the infinite-size limit $N\rightarrow\infty$ coincides with the classical limit $\hbar_{\mathrm{eff}}\rightarrow0$ in this model.

In the classical limit, the operators $\hat{D}_{x,y}$ and $\hat{R}_{x,y}$ behave approximately as a momentum and a position coordinate in the phase space~\cite{SanchezCastellanos2009,EstevezFregoso2018} as they are mapped to 
\begin{subequations}
\label{eq:Dclassical}
\begin{align}
    \frac{1}{N}\hat{D}_{x,y} &\xrightarrow{N \rightarrow \infty} p_{2,1}\sqrt{2 - s^2},\\
    \frac{1}{N}\hat{R}_{x,y} &\xrightarrow{N \rightarrow \infty} q_{1,2}\sqrt{2 - s^2}.
\end{align}
\end{subequations}
Due to the above-mentioned connection of the $\overline{\alg{o}}(3)$ and $\alg{o}(3)$ operators via a unitary transformation, the spectrum of the Hamiltonian is the same no matter whether constructed of the $\hat{D}_{x,y}$ or $\hat{R}_{x,y}$ operators.

\section{Results}
\label{sec:Results}
In this section, we begin by illustrating the manifestation of classical chaos in the classical limit of the $\alg{u}(3)$ model. Then we briefly explain numerical techniques used to calculate the Lyapunov exponent $\lyapunov$ and the fraction of regularity $\freg$.
We fit the short-time evolution of the OTOC with the exponential function~\eqref{eq:InitialGrowth} and confirm that its exponential rate $\lambda$ corresponds with the classical Lyapunov exponent for various OTOC operators.
Finally, we present the main topic of this paper---the OTOC fluctuations at long times.
All the calculations were performed in the programming language Julia.
The classical dynamics was computed using the package DifferentialEquations.jl~\cite{Rackauckas2017}.

\subsection{Classical chaos}
\begin{figure}[!htbp]
    \centering
    \includegraphics[width=\linewidth]{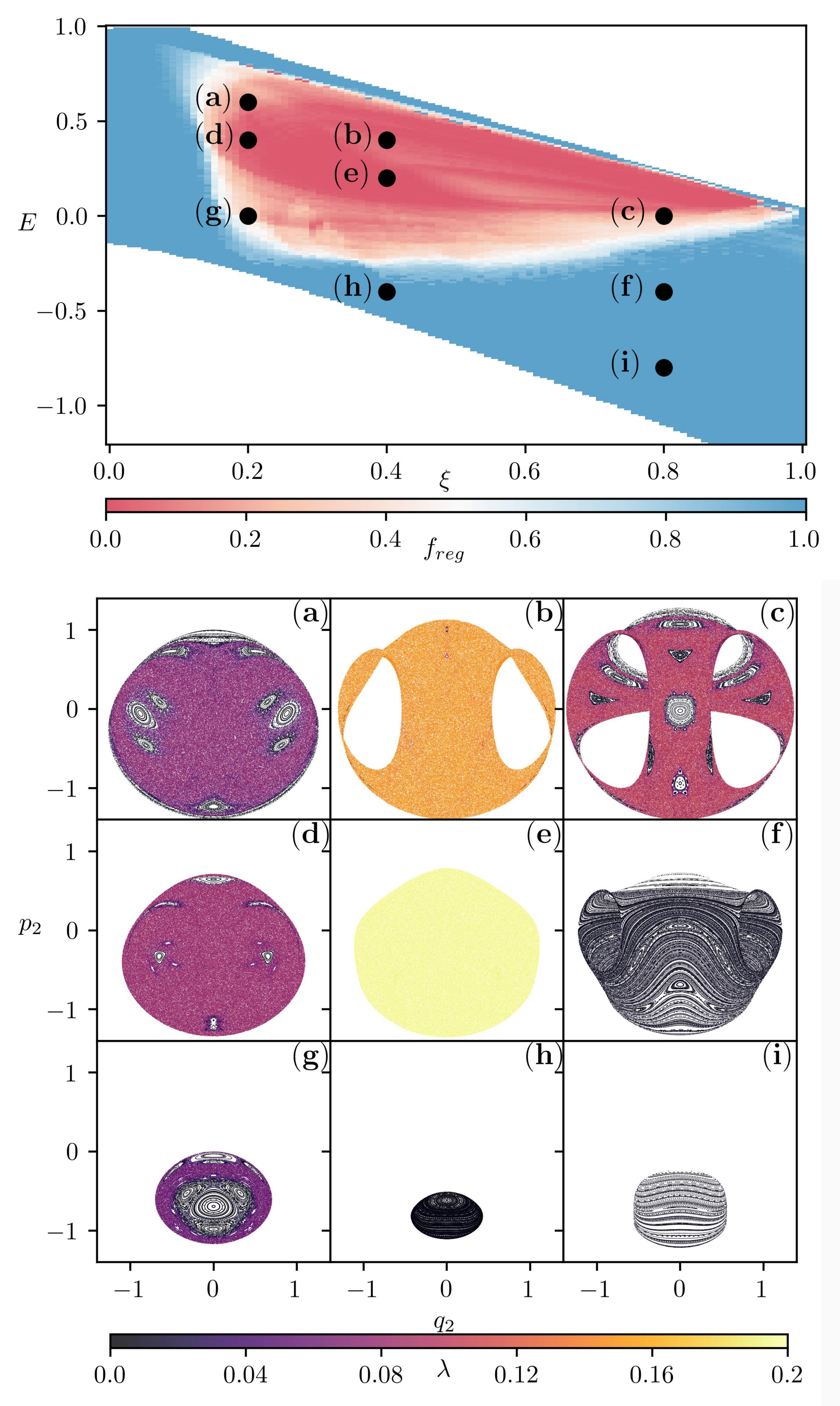}
    \caption{
        Classical fraction of regularity $\freg(\xi,E)$ (top panel) and selected examples of Poincaré sections by $q_{1}=0$ plane [panels (a)---(i)] for the classical limit of the $\alg{u}(3)$ model~\eqref{eq:hcl} with nonintegrable perturbation $\epsilon=0.4$.
        The colors of the section points indicate the values of the Lyapunov exponent $\lambda_{\mathrm{cl}}$ for the corresponding trajectories.
        Values $\freg=0$ and $\freg=1$ correspond to fully chaotic (all trajectories on the energy hypersurface $\Sigma_E$ have a positive Lyapunov exponent) and fully regular (all trajectories have $\lyapunov=0$) dynamics, respectively.
    }    
    \label{fig:fregPoincare}
\end{figure}

Except for some particular values of the parameters described in Sec.~\ref{sec:Model}, the model described by Hamiltonian~\eqref{eq:hcl} is nonintegrable, thus exhibiting chaotic dynamics.
In Fig.~\ref{fig:fregPoincare}, we show the classical fraction of regularity $\freg$ for a fixed value of external perturbation $\epsilon = 0.4$ and for $\xi \in [0,1]$. 
The chaotic region ($\freg < 1$) starts to arise at $\xi \approx 0.1$ and lies approximately on and above the energy $E = 0$.
The lowest and highest energy limits of the system are fully regular because the classical Hamiltonian function~\eqref{eq:hcl} is approximately quadratic near its global minimum and maximum, leading to stable dynamics.
The chaotic region has an intricate structure with islands of enhanced regularity embedded in areas of almost complete chaos, caused by the appearance and disappearance of regular tori~\cite{Stransky2009} and resulting in sudden peaks and dips in curves $\freg(E)$ for fixed values of $\xi$, see Fig.~\ref{fig:FregVSFanoVSlyap}.

The chaoticity is further demonstrated by Poincaré sections plotted in panels (a)---(i). Each section point corresponds to one trajectory crossing through the chosen section plane $q_{1}=0$; the color of the points indicates the size of the Lyapunov exponent. Stable trajectories form regular structures in the shape of deformed lines or circles, whereas chaotic areas are filled with seemingly randomly located points.
In general, these 2D Poincaré sections visualize the 3D energy manifolds of constant energy embedded in the 4D phase space. The white areas in panels (b) and (c) are proper topological holes in the energy hypersurface and depict the intriguing shape of the sections.

The values of the Lyapunov exponents for individual trajectories were obtained by evolving the deviation $\delta\mathbf{x}$ of two close orbits for a sufficiently long time using the equation of tangent dynamics~\cite{Skokos2010,Lerose2020,PilatowskyCameo2020}. 
As the $\freg(E)$, we effectively take the ratio of the regular area (area in the Poincaré section $q_{2}=0$ crossed by trajectories with $\lyapunov=0$) to the total section area intersected by any of the orbits. We choose the initial conditions so that the given Poincaré section divided into a sufficiently dense mesh of cells is entirely covered with evolved trajectories. 
The energy Lyapunov exponent $\lyapunove(E)$ is calculated in the same manner as an average of Lyapunov exponents of each cell in the mesh.

Our detailed numerical analysis shows that the values of $\freg$ obtained from the Poincaré sections differ from the $\freg$ computed directly from~\eqref{eqn:f_regDefinition} (\emph{i.e.} as a ratio of the volumes of the 3D hypersurfaces) by no more than $\approx 5\%$, which is a fair price to pay for saving plenty of the computation time. Note that the same procedure of calculating the $\freg$ was used and discussed in another model~\cite{Stransky2009}.  

\subsection{Short-time OTOCs}

\begin{figure}[tbp]
    \centering
    \includegraphics[width=\linewidth]{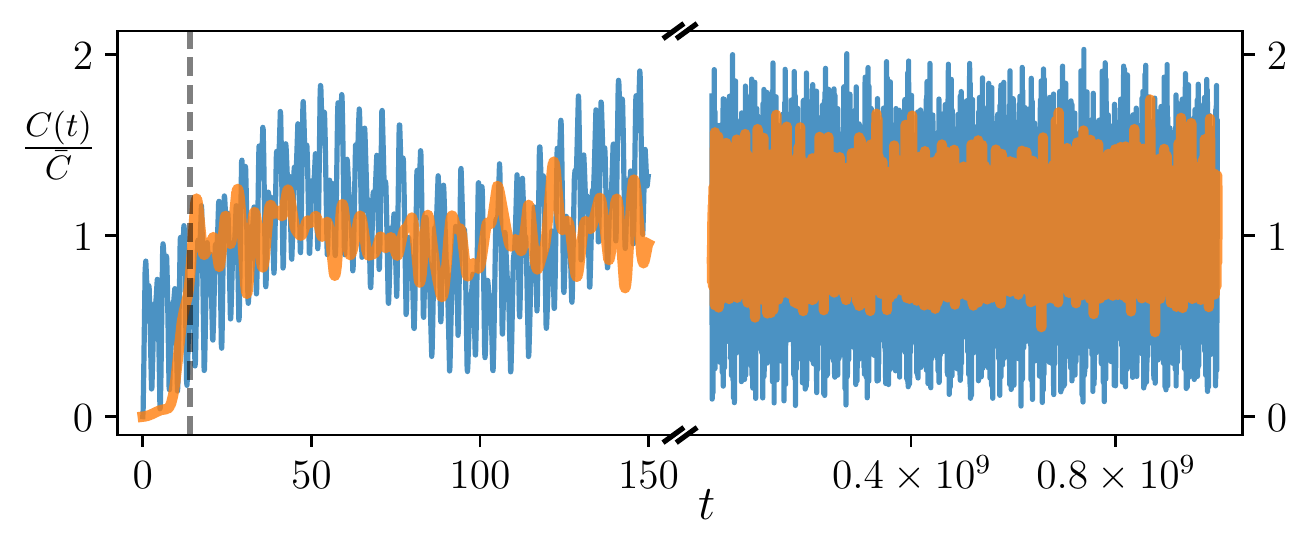}
    \caption{
        An illustration of the short-time and long-time OTOC dynamics for operators $[\hat{D}_x(t),\hat{D}_x(0)]^2$ in a state from a fully regular ($n=1$, thin blue curve) and chaotic ($n=650$, thick orange curve) part of the quantum energy spectrum of the $\alg{u}(3)$ model, respectively, as a function of time. Model parameters are $\xi = 0.4$, $\epsilon = 0.4$, system size is $N = 50$. OTOC values are divided by the corresponding mean value $\overline{C}$ for better comparison; $\overline{C}(E_{1}=-0.54) \approx 10^1$ and $\overline{C}(E_{650} = 0.21) \approx 10^5$.   
        The vertical black dashed line indicates approximate Ehrenfest time $\widetilde{t}$ for the chaotic state.
    }
    \label{fig:OTOC}
\end{figure}

As explained in Sec.~\ref{sec:CQC}, the short-time behavior of the OTOCs reflects the classical divergence of neighboring trajectories and can serve as a quantum analog of the classical Lyapunov exponent.
In Fig.~\ref{fig:OTOC} we show the typical time series of the OTOCs for two different quantum states, one taken from the region of classically regular dynamics and the other from the classically chaotic region (the states will be called regular and chaotic for brevity).
The regular state is characterized by wide oscillations, with the OTOC values frequently returning close to the initial value $C(0)$. In contrast, the chaotic state starts with a fast (exponential) increase followed by small oscillations around the saturation value.

\begin{figure}[tbp]
    \centering
    \includegraphics[width=\linewidth]{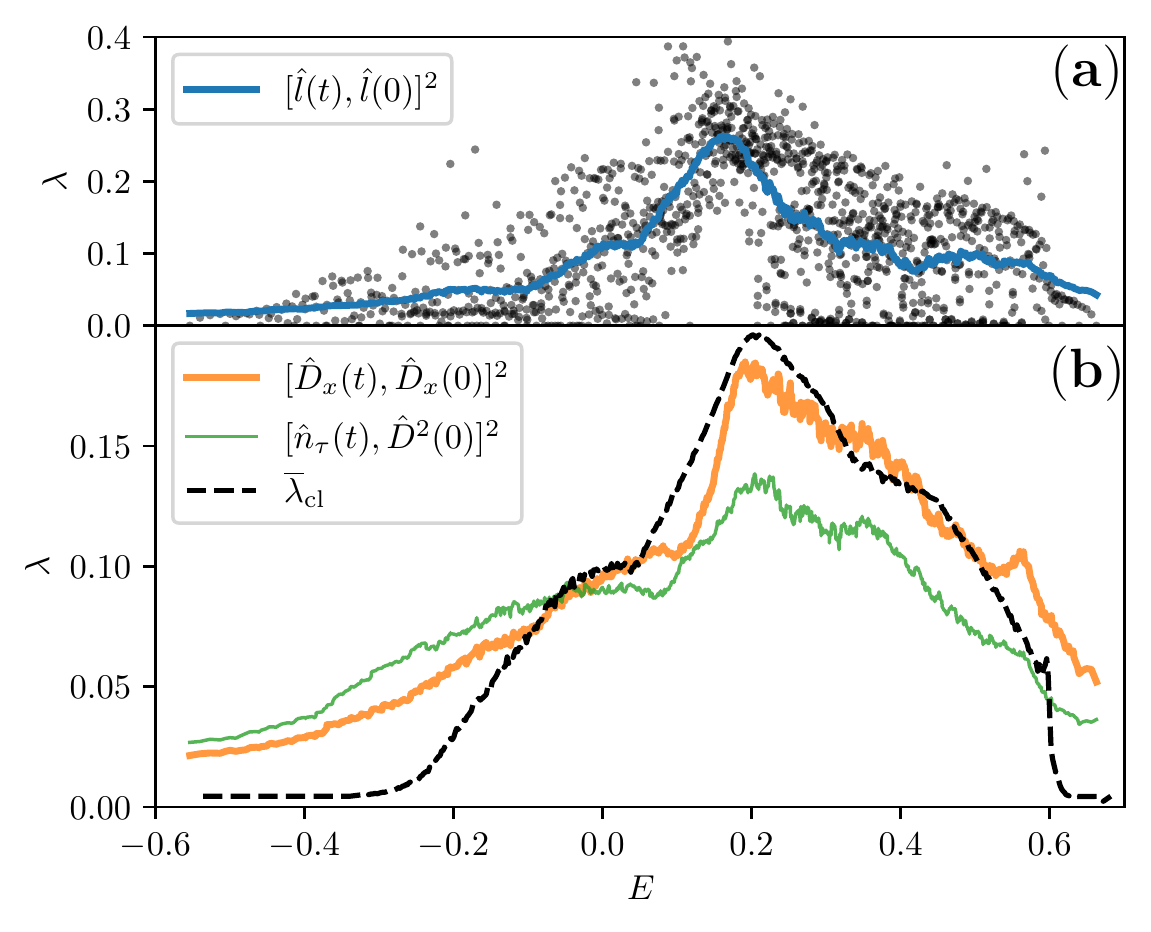}
    \caption{(a) Instability parameters $\lambda_{n}$ from the exponential fit of the short-time growth of OTOC $[\hat{l}(t), \hat{l}(0)]^{2}$ for all energy eigenstates (black dots), and a smoothed value by a moving average over $60$ consecutive states (blue curve).
    (b) Smoothed instability parameters $\lambda$ for two other OTOC operators indicated in the legend (solid lines), compared with the classical energy Lyapunov exponent $\lyapunove$ (dashed line). Parameters of the $\alg{u}(3)$ model are $\xi=0.4$, $\epsilon=0.4$. System size is $N=50$.
    }
    \label{fig:Lyap}
\end{figure}

The connection between the short-time OTOC behavior and classical chaos is demonstrated in Fig.~\ref{fig:Lyap}. 
For each state, we calculate the instability parameter $\lambda_{n}$ from the exponential fit of the initial OTOC growth~\eqref{eq:InitialGrowth}, see the black dots in Fig.~\ref{fig:Lyap}(a).
As the final time of the fit we take the smallest $\widetilde{t}_{n}$ satisfying equation $C_{n}(\widetilde{t}_{n})=\overline{C}_{n}-\sigma_{n}$ (for chaotic states, $\widetilde{t}_{n}$ can also serve as an estimate of the Ehrenfest time $t_{E}$).
Then, we take a moving average over a window of $60$ neighboring states to obtain a smooth curve.
The smooth curves for two other OTOC operators are compared with the classical Lyapunov exponent in Fig.~\ref{fig:Lyap}(b).
Despite the roughness of the whole procedure, a qualitative agreement between the computed energy dependence of the quantum instability parameter for all OTOC operators and $\lambda(E)$ is observed: An initial increase from low values of $\lambda$ reflecting regular dynamics close to the ground state of the system, followed by a plateau at $E\approx0$ and a pronounced global maximum at $E\approx0.2$.
The stability is enhanced again at the upper edge of the spectrum.

Higher system size $N$ leads to (i) a denser quantum spectrum, allowing for a wider moving average window with maintaining the desired energy resolution, and (ii) longer Ehrenfest time $t_{E}$, resulting in better fits of $\lambda_{n}$. Therefore, it further improves the agreement between classical and quantum Lyapunov exponents.

\subsection{Long-time OTOCs}
The focus of this paper is the analysis of the OTOCs for energy eigenstates at asymptotic times.
A brief glimpse at Fig.~\ref{fig:OTOC} reveals the qualitative difference between the relative OTOC oscillations in the regular and chaotic state at higher times. The OTOC function evaluated for the regular state with energy $E_{1}$ forms a wide ``strip'' covering the graph from $C\approx 0$ to some finite value $C=C_{\mathrm{max}}$, usually, but not always, resulting in a relatively high value of the wiggliness. In contrast, the chaotic state, after the initial exponential explosion, covers a narrower strip with minor oscillations around the saturation value, and the wiggliness tends to be smaller.

\begin{figure*}[tbp]
    \centering
    \includegraphics[width=\linewidth]{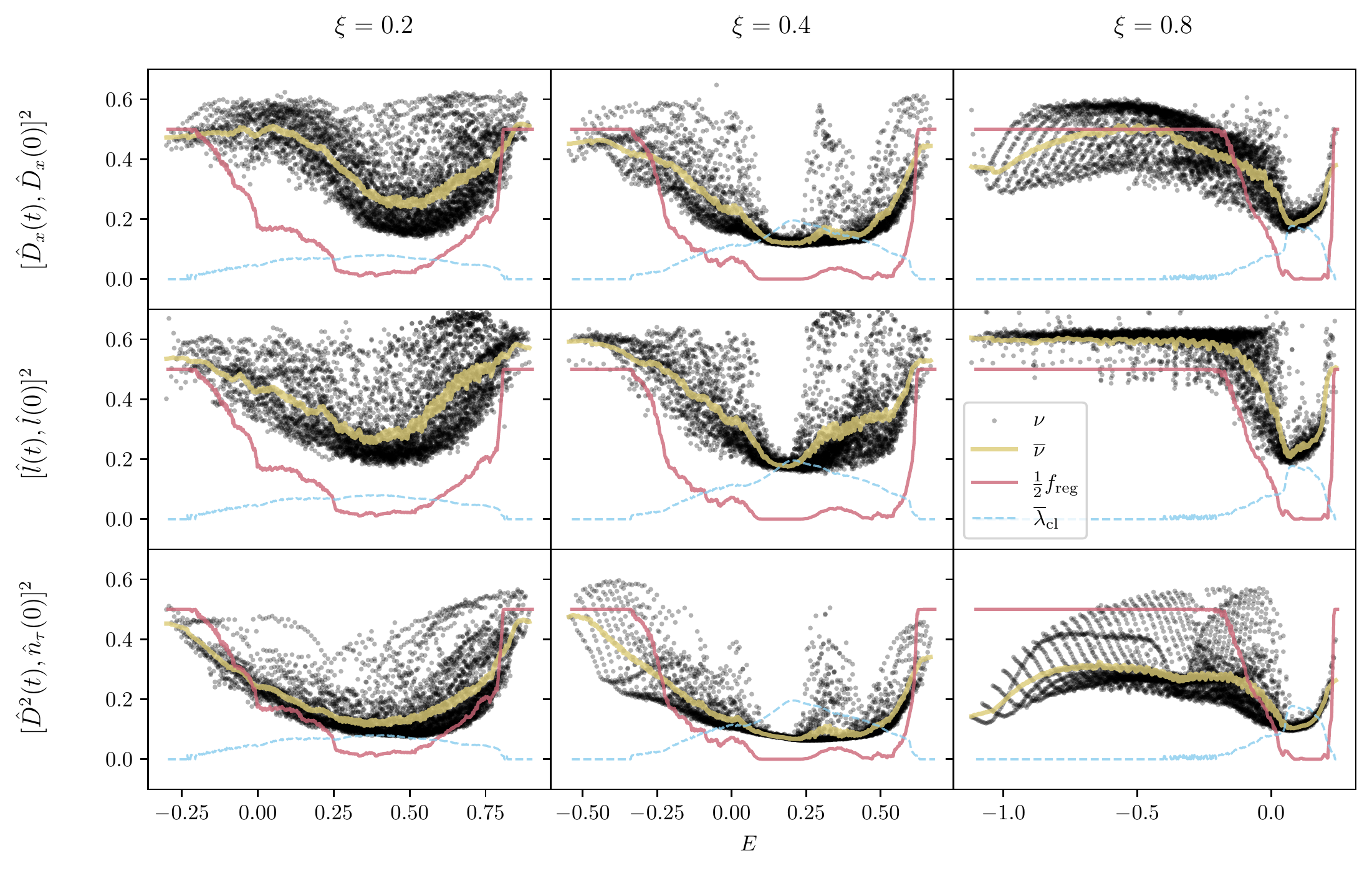}

    \caption{
        Wiggliness for three distinct pairs of the OTOC operators calculated for all eigenstates $|E_{n}\rangle$ along the spectrum (black dots), classical fraction of regularity $\freg$ (thin solid red curve) and the classical energy Lyapunov exponent (dashed blue curve). In addition, the wiggliness is smoothed using the moving average over $50$ consecutive states (thick solid yellow curve). The OTOC operators are indicated at the vertical axes. Parameters of the $\alg{u}(3)$ model are $\xi = \{0.2,0.4,0.8\}$ (in columns), $\epsilon = 0.4$. System size is $N = 100$. Note that the Poincaré sections in Fig.~\ref{fig:fregPoincare} were computed for the same model parameters and can serve as an additional illustration of the regular versus chaotic dynamics.
    }
    \label{fig:FregVSFanoVSlyap}
\end{figure*}

The wiggliness for all energy eigenstates in several system configurations is shown by black dots in Fig.~\ref{fig:FregVSFanoVSlyap}.
Each column corresponds to a different choice of parameter $\xi$ of the model, hence different chaotic properties, whereas the rows show the wiggliness for various OTOC operators.
The classical fraction of regularity $\freg$ is drawn by a medium-thick red curve.
In this and all subsequent figures, the mean values~\eqref{eq:OTOCmean} and variances~\eqref{eq:OTOCvar} for the wiggliness are calculated from the OTOCs in a random sample of $2500$ time values uniformly distributed over the interval $10^{7}<t<10^{9}$ of sufficiently high times, much greater than the Ehrenfest time.

We distinguish three types of wiggliness behavior based on the chaoticity of the system:
\begin{itemize}
    \item {\textbf{Regular regions} ($\freg = 1$)}

    In the parts of the energy spectra that correspond to completely regular dynamics, the values $\nu_{n}$ are higher in comparison with other regions, reaching usually values around $\nu\approx0.6$ independently of the system size, as will be seen later. They may form regular structures as $\nu_{n}$ reflects the existence of an additional constant of motion, see Fig.~\ref{fig:FregVSFanoVSlyap}~(c),(h),(i). The shape of these structures naturally depends on the choice of OTOC operators.
    
    \item {\textbf{Chaotic regions} ($\freg = 0$)}
    
    For quantum states that lie in completely chaotic energy intervals, the values $\nu_{n}$ are generally smaller compared to other regions. In addition, the wiggliness values corresponding to neighboring eigenstates are much more similar. This can be observed in Fig.~\ref{fig:FregVSFanoVSlyap} for $\xi = 0.4$ (second column) at $E\approx0.2$ and $\xi = 0.8$ (third column) at $E\approx0.1$, as all the points in these chaotic regions are ``pressed'' together.
    
    \item {\textbf{Mixed spectrum} ($\freg > 0$)}
    
    In the partially chaotic energy intervals of the spectra, there are intertwined sets of states with properties of both regular and chaotic regions. Therefore, they are characterized by a relatively wide spread of wiggliness values but without any characteristic regular pattern. 
    The sensitivity of $\nu_{n}$ to the presence of classical regular islands is remarkable. Consider the region $0.25<E<0.5$ in the second column of Fig.~\ref{fig:FregVSFanoVSlyap}. The system is almost entirely chaotic at this energy interval in the classical limit. In the corresponding Poincaré section shown Fig.~\ref{fig:fregPoincare}(b), we can hardly notice tiny regular islands immersed in the chaotic sea. Yet there is a significant amount of eigenstates with high OTOC wiggliness, reflecting the tiny spots of classical regularity.
\end{itemize}

To see the general trend, we calculated the smoothed value $\overline{\nu}(E)$ of the wiggliness as a moving average over a small number of successive individual energy levels and showed in in Fig.~\ref{fig:FregVSFanoVSlyap} (thick yellow curve).
The smoothed value reveals that the average wiggliness is strongly correlated with the fraction of regularity $\freg$.
It leads us to conjecture that the smoothed asymptotic relative oscillations of the eigenenergy OTOCs reflect the global chaoticity of the quantum system at the smoothing energy window.

The structures present for the wiggliness in the regular regions of energy can be understood within the theory of Peres lattices~\cite{Peres1984,Stransky2009a}, formed from energy expectation values $O_{n}=\langle E_{n}|\hat{O} |E_{n}\rangle$ of an operator $\hat{O}$ as a set of points $(E_n, O_n)$.
In nonchaotic parts of the energy spectrum, the observable $\hat{O}$ can be expressed as a smooth function of all the independent constants of motion, which leads to regular patterns in the Peres lattice.
Chaos appears with the loss of integrals of motion, which leads to a collapse of the regular Peres lattice. In the fully chaotic regions, all the points are ``pressed'' together with minor differences in the $O_{n}$ expectation values. The differences further diminish with the increase in the system size. This behavior can be understood via the Eigenvector Thermalization Hypothesis~\cite{Deutsch1991,Srednicki1994,Rigol2008}. 
Values $\overline{C}_{n}$ and $\sigma_{n}$ given by Eqs.~\eqref{eq:OTOCmean} and~\eqref{eq:OTOCvar} are, in fact, expectation values of rather complicated time-averaged operators provided by the OTOC commutators, and the wiggliness is a ratio of these two; hence it is expected that the wiggliness itself retains some of the properties of the Peres lattices.

\begin{figure}[tbp]
    \centering
    \includegraphics[width=\linewidth]{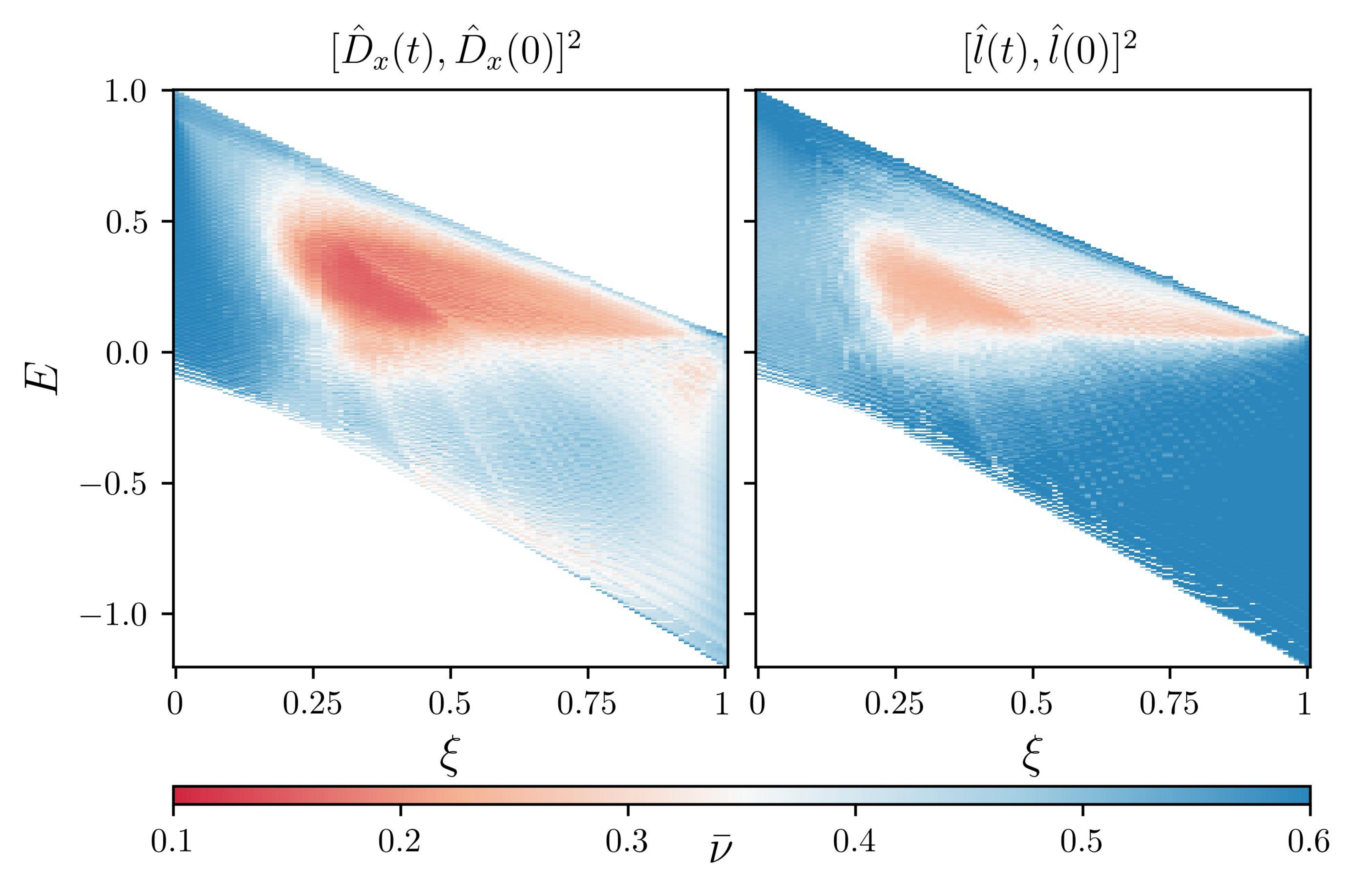}
    \caption{Smoothed wiggliness $\overline{\nu}$ along the spectra of the $\alg{u}(3)$ model~\eqref{eq:hamiltonian} for $\xi \in [0,1]$, $\epsilon = 0.4$ and two pairs of OTOC operators. The smoothing window entails five neighboring levels. The system size is $N = 60$, corresponding to 1891 energy levels. Roman numbers I and II label two regions with spuriously high values of $\overline{\nu}$ for OTOC operator $[\hat{D}_x(t),\hat{D}_{x}(0)]^{2}$ (see the main text).}
    \label{fig:FanoHeatmap}
\end{figure}

Fig.~\ref{fig:FanoHeatmap} displays the smoothed wiggliness for two choices of OTOC operators
$[\hat{D}_x(t),\hat{D}_x(0)]^2$ and $[\hat{l}(t),\hat{l}(0)]^2$. We can directly compare these results with the $\freg$ plotted in Fig.~\ref{fig:fregPoincare} and observe that the chaotic regions with $\freg\approx0$ and regions with reduced $\overline{\nu}(E)$ overlap. In the case of the operator $\hat{D}_x$, we find regions of small smoothed wiggliness in the regular area of the spectrum (marked I and II in the figure), spuriously labeling them as chaotic. These regions of suppressed wiggliness can also be seen in Fig.~\ref{fig:FregVSFanoVSlyap}(i). Low values of $\overline{\nu}(E)$ in region I are related to the fact that $\xi=0.8$ is very close to the value $\xi=1$ for which $\hat{D}_{x}$ commutes with the Hamiltonian. Region II is a finite-size effect that diminishes with increasing $N$. 

\begin{figure}[tbp]
    \centering
    \includegraphics[width=\linewidth]{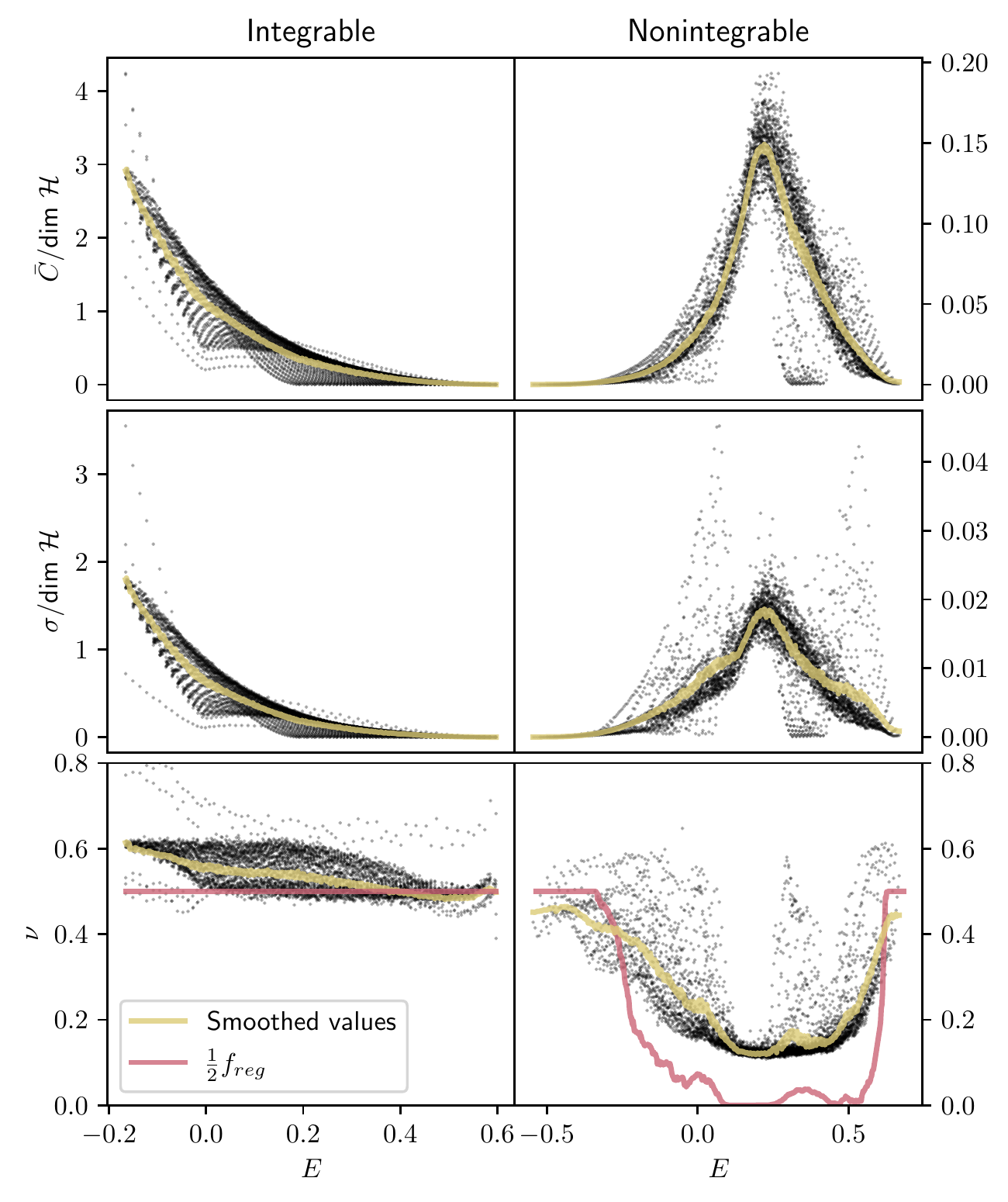}
    \caption{Comparison of mean values $\overline{C}_{n}$ (first row), standard deviations $\sigma_{n}$ (second row) and the wiggliness $\nu_{n}$ (third row) of OTOC $[\op{D}_x(t),\op{D}_x(0)]^2$ in an integrable case $\epsilon = 0$ (left column) and chaotic case $\epsilon = 0.4$ (right column) of the $\alg{u}(3)$ model with $\xi = 0.4$ and system size $N = 100$. Values of $\overline{C}_{n}$ and $\sigma_{n}$ are divided by the Hilbert space dimension $\dim\mathcal{H}= 5151$. Thick yellow lines correspond to smoothed values using the moving average over $50$ consecutive states. The thin red line is the fraction regularity $\freg$, scaled by 1/2 for better visualization.}
    \label{fig:VarAndMean}
\end{figure}

So far, we have discussed the wiggliness only. However, one could ask whether the mean value and variance on their own (or the standard deviation as the square root of the variance) would be sufficient to indicate and quantify the presence of chaos. 
Based on extensive numerical evidence, the answer is negative. 
This is illustrated in Fig.~\ref{fig:VarAndMean} where we compare $\overline{C}_{n}$, $\sigma_{n}$ and $\nu_{n}$ of the OTOC of $[\op{D}_x(t),\op{D}_x(0)]^2$ in an integrable and nonintegrable regimes of the $\alg{u}(3)$ model. As mentioned above, the sets of points $(E_{n},\overline{C}_{n})$ and $(E_{n},\sigma_{n})$ form a kind of Peres lattices. Therefore we can infer the presence of chaos in energy regions where the lattice is disordered, and expect full chaoticity where the disordered lattice shrinks into a vertically narrow band. In the nonintegrable regime, the mean value and the standard deviation are higher in the chaotic regions around $E\approx0.25$ than in the regular regions close to the lower and upper energy limits of the spectrum. However, their values are more than an order of magnitude lower than in the integrable case $\epsilon = 0$. Therefore we cannot simply determine the chaoticity of a general system based on the concrete values of $\overline{C}_{n}$ and $\sigma_{n}$, even after smoothening over several consecutive states. It is their ratio in the form of relative oscillations that reveals the degree of chaoticity and that corresponds to the classical $\freg$, as demonstrated in Figs~\ref{fig:FregVSFanoVSlyap},~\ref{fig:FanoHeatmap} and ~\ref{fig:VarAndMean}(c).

\subsection{Effect of the system size}

\begin{figure}[tbp]
    \centering
    \includegraphics[width=1\linewidth]{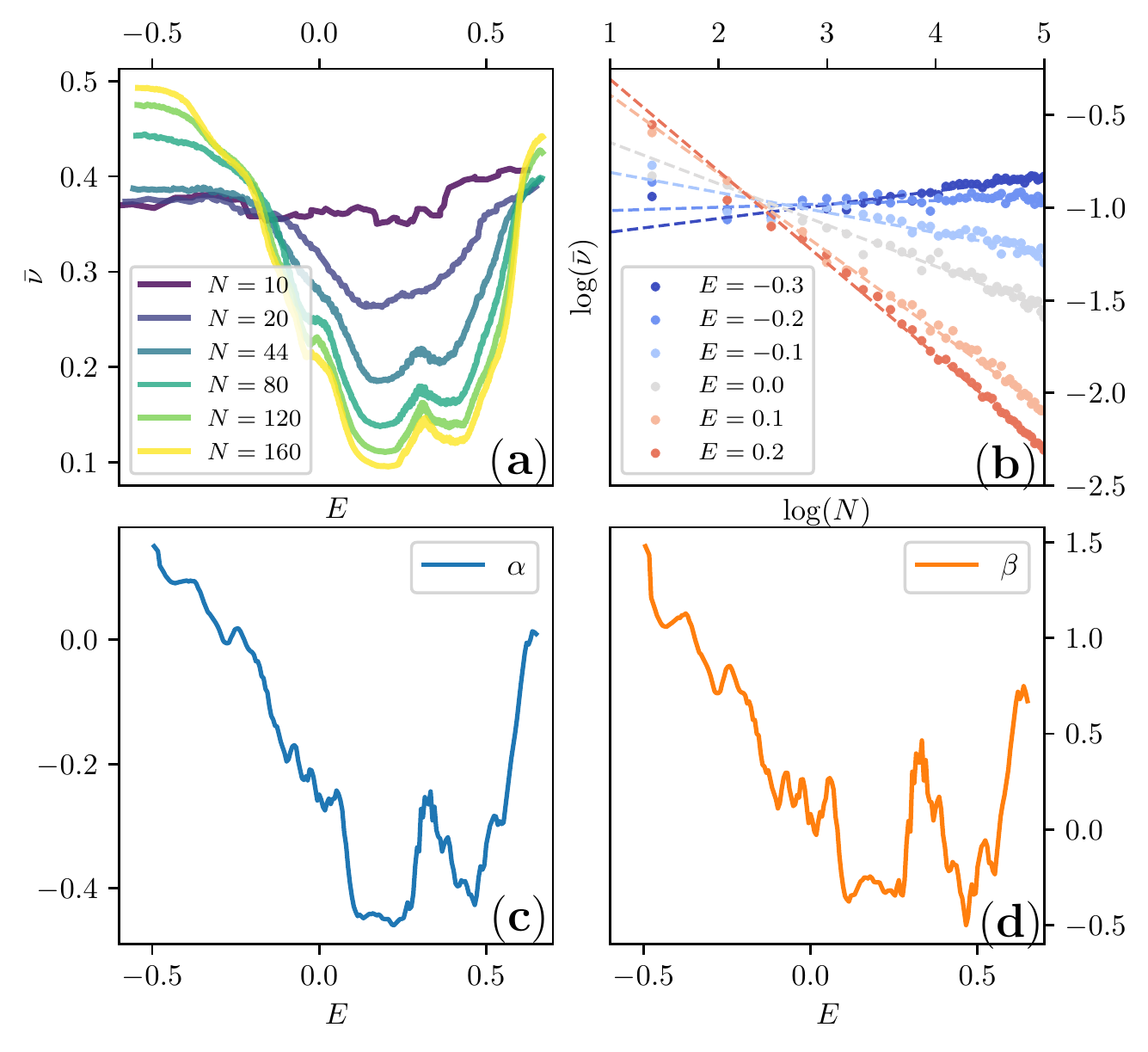}
    \caption{Dependence of the wiggliness on the size $N$ of the quantum system. Figures are calculated for the $\alg{u}(3)$ model in the chaotic regime with $\xi = 0.4$ and $\epsilon = 0.4$. As the OTOC operators we choose $[\hat{D}_x(t),\hat{D}_x(0)]^2$. Panel (a) displays smoothed wiggliness (the width of the smoothing window is taken as $2N$ here) as a function of energy along the spectrum for several system sizes ranging from $N = 10$ to $N = 160$, as indicated in the legend. Panel (b) shows the dependence of the smoothed wiggliness on $N$ for several energy values in the log-log scale (points) and their fit by function~\eqref{eq:Fano-N,E} (dashed lines).
    Panels (c) and (d) show the coefficients $\alpha$ and $\beta$, respectively, along the whole energy range of the spectrum.}
    \label{fig:OTOCsN}
\end{figure}
    
Here we discuss the effect of the system size on the relative oscillations $\nu_{n}$. We take the total number of boson excitations $N$ of the $\alg{u}(3)$ model as a variable and study the dependence of the smoothed wiggliness $\overline{\nu}(N)$ both on the energy $E$ and on the size parameter $N$.

Fig.~\ref{fig:OTOCsN}(a) shows an example of the smoothed wiggliness for one choice of the pairs of OTOC operators for the whole accessible energy range and several values of $N$. The observed overall behavior is the following: In the regions of the spectrum with chaotic or mixed dynamics ($\freg<1$), $\overline{\nu}$ decreases with increasing size of the system, whereas in the regular regions ($\freg=1$) the smoothed wiggliness remains approximately constant (usually with a value $\approx0.6$).
This observation is further confirmed in Fig.~\ref{fig:OTOCsN}(b), which shows the dependence $\overline{\nu}(N)$ for six different values of energy taken from regions of both regular and chaotic dynamics. The linear dependence in the log-log plot implies algebraic scaling of $\overline{\nu}$ with $N$,
\begin{equation}\label{eq:Fano-N,E}
\overline{\nu}(E, N)=N^{\alpha(E)} e^{-\beta(E)},
\end{equation}
where the dependence of $\overline{\nu}$ on the energy manifests only via the exponents $\alpha(E)$ and $\beta(E)$. These exponents, obtained by the fit as demonstrated in Fig.~\ref{fig:OTOCsN}(b), are plotted in Figs.~\ref{fig:OTOCsN}(c) and (d).
Since the spectrum of the system is discrete, we interpolated the $\overline{\nu}$ values to get precisely into the desired energy value for the fit.

In Figs.~\ref{fig:OTOCsN}(c) and (d), one can immediately notice the apparent qualitative resemblance of $\alpha(E)$, $\beta(E)$, and the wiggliness itself, see Figs.~\ref{fig:FregVSFanoVSlyap}(b) and~\ref{fig:OTOCsN}(a). 
The most important is the correspondence between $\overline{\nu}(E)$ and the scaling exponent $\alpha(E)$. It implies that the suppression rate of the wiggliness scales with the system size; the more chaotic the system is, the faster the suppression of $\overline{\nu}$.
Focusing on the wiggliness of individual states gives a clue for this behavior: While the wiggliness of the regular states remains constant with changing system size, it diminishes for chaotic states. The more states categorized as chaotic in regions with mixed dynamics, the faster $\alpha(N)$ decreases.
The exponent $\alpha$ can, therefore, also serve as a chaos indicator, and because it does not depend on the size $N$ of the system, it can be considered even more robust than the wiggliness itself. 
Its behavior can be summarized in the following way: In energy regions of the spectrum with chaotic or mixed dynamics ($\freg<1$), the suppression rate is negative, $\alpha<0$, whereas in regular regions ($\freg=1$) there is no suppression of the wiggliness, $\alpha\gtrsim0$.
Note that the scaling of the OTOC relative oscillations has already been observed for non-collective spin systems~\cite{Anand2021}.

Apart from the correspondence between the scaling exponent $\alpha$ and the smooth wiggliness $\overline{\nu}$, there is also an evident similarity between $\alpha(E)$ and $\beta(E)$, see Figs.~\ref{fig:OTOCsN}(c) and (d).
This resemblance is caused by the fact that in the log-log representation, Eq.~\ref{eq:Fano-N,E} is a linear function, $\log\overline{\nu}=\alpha\log{N}-\beta$. If the lines corresponding to different values of $\alpha$ intersect in the same quadrant of the graph, as seen in Fig.~\ref{fig:OTOCsN}(b), then the intercept given by $\beta$ will follow the same trend as the slope $\alpha$.

\begin{figure}[tbp]
    \centering
    \includegraphics[width=1.\linewidth]{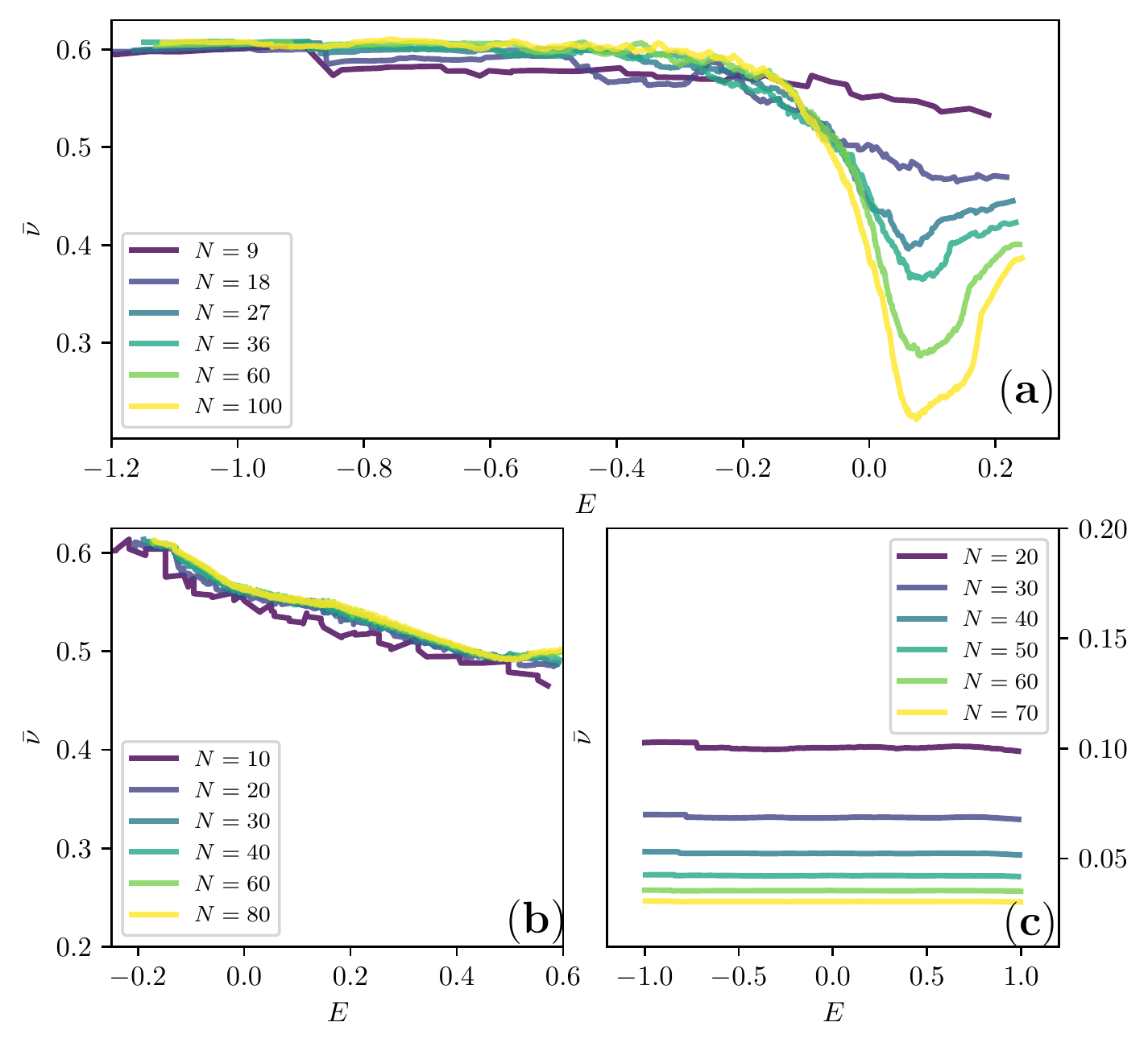}
    \caption{Smoothed wiggliness $\overline{\nu}$ plotted for various system sizes $N$ and in three different models. (a) $\alg{u}(3)$ model with $\xi = 0.8$, $\epsilon = 0.4$ and OTOC operators $[\hat{l}(t),\hat{l}(0)]^2$. (b) $\alg{u}(3)$ model with $\xi = 0.4$, $\epsilon = 0.0$ (integrable case) and OTOC operators  $[\hat{D}_x(t),\hat{D}_x(0)]^2$. (c) A fully chaotic system with Hamiltonian modeled by a random GOE matrix~\cite{Haake2010} and OTOC operators $[\hat{D}_x(t),\hat{D}_x(0)]^2$ taken from the $\alg{u}(3)$ model.}
    \label{fig:OTOCsNmulti}
\end{figure}

\begin{figure}[tbp]
    \centering
    \includegraphics[width=\linewidth]{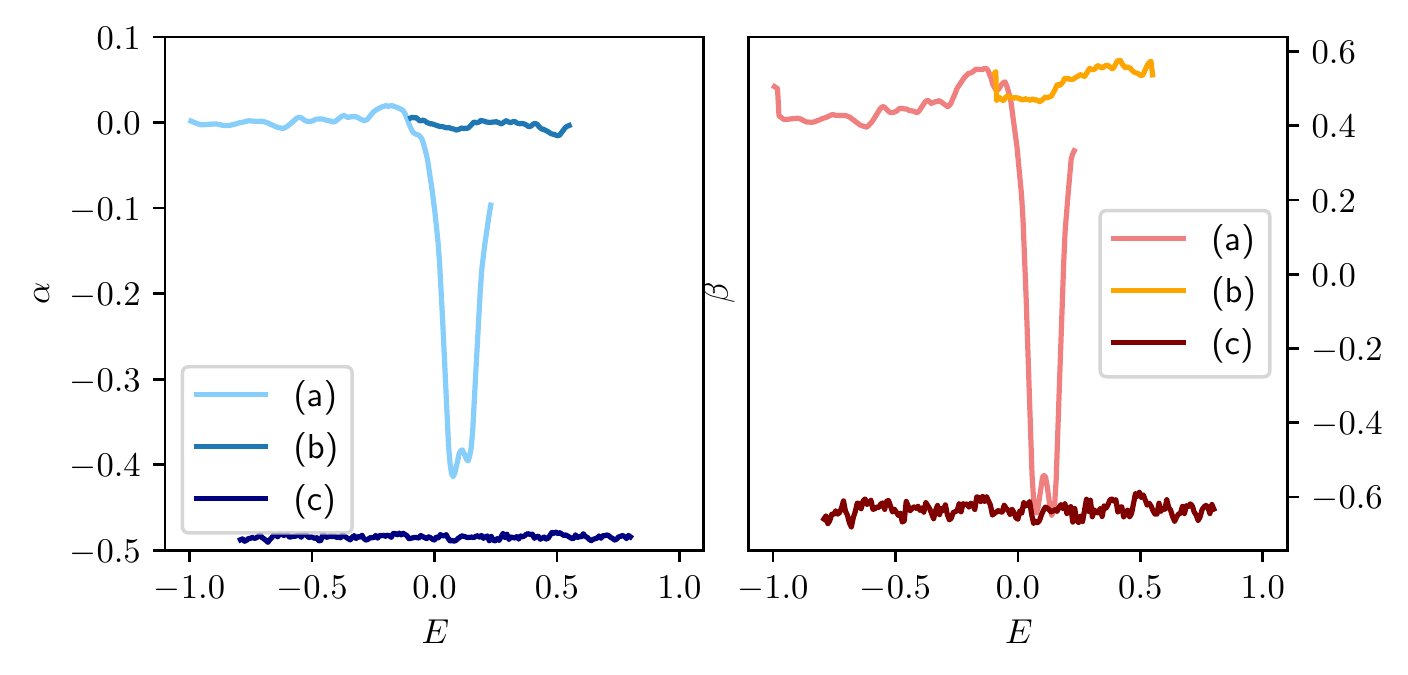}
    \caption{Parameters $\alpha$ and $\beta$ obtained for the three models (a), (b), (c) displayed in Fig.~\ref{fig:OTOCsNmulti}.}
    \label{fig:OTOCsNmultiAB}
\end{figure}

We finish this section by demonstrating the validity of the scaling conjecture for three different regimes: (a) another case of the mixed dynamics in the $\alg{u}(3)$ model [Fig.~\ref{fig:OTOCsNmulti}(a)], (b) integrable, hence fully regular dynamics in the $\alg{u}(3)$ model [Fig.~\ref{fig:OTOCsNmulti}(b)] and (c) entirely chaotic dynamics simulated by an artificial system with Hamiltonian taken as a random GOE matrix~\cite{Haake2010} [Fig.~\ref{fig:OTOCsNmulti}(c)], whose dimension corresponds to the dimension of the $\alg{u}(3)$ model~\eqref{eq:dimension} with size $N$. As before, we observe that the wiggliness is algebraically suppressed in the chaotic regions of spectra and approximately constant in the regular parts with increasing system size $N$. Corresponding exponents $\alpha$ and $\beta$ are shown in Fig.~\ref{fig:OTOCsNmultiAB}. Regular parts of spectra have almost zero suppression rate $\alpha$. In contrast, regions with mixed or purely chaotic dynamics have negative $\alpha$, which also mimics the shape of $\freg(E)$ and reflects the level of chaoticity. 

In case (c) of the GOE Hamiltonian, the scaling exponent is negative with a value $\alpha\approx-1$ independent of a choice of the eigenstate. 
It is consistent with the fact that the GOE matrix models a fully chaotic case. However, the parameter $\alpha$ is smaller than the lowest values of $\alpha$ observed in the fully chaotic energy intervals of the $\alg{u}(3)$ model; see Fig.~\ref{fig:OTOCsNmultiAB} for line (a) at $E\approx0.1$ or Fig.~\ref{fig:OTOCsN}(c) at $E\approx0.1$ and compare it with corresponding $\freg$ curves in Fig.~\ref{fig:FregVSFanoVSlyap}.
This discrepancy is caused by a partial localization of the $\alg{u}(3)$ eigenvectors, even in the fully chaotic energy regions.
On the other hand, the GOE eigenstates are, by construction, maximally delocalized.
Thus the GOE system gives a lower limit for the decay speed of the wiggliness, but its $\alpha$ cannot serve as a reference value for general chaotic quantum systems.

Note that the same scaling behavior of the wiggliness is observed for other choices of OTOC operators and parameters of the algebraic $\alg{u}(3)$ model.

\section{Conclusions}\label{sec:Conclusions}
We have demonstrated that the long-time OTOC dynamics in energy eigenstates account for the overall chaoticity of the quantum system, in contrast with the short-time OTOC behavior that testifies about the stability of quantum states and is used as the quantum analog of the Lyapunov exponent.
The quantum Lyapunov exponent thus reflects local properties of the dynamics, while the dimensionless asymptotic relative oscillations---the wiggliness---are related to the global characteristics of chaoticity.
It is analogical to classical mechanics, where the classical Lyapunov exponent characterizes a single trajectory, whereas an asymptotic time evolution of a chaotic orbit, due to ergodic and topological properties, covers the chaotic domain of the phase space energy hypersurface.
Then, from the knowledge of the entire volume of the energy hypersurface, one can immediately deduce the fraction of regularity $\freg$.

By an extensive study of various OTOCs in different regimes of the $\alg{u}(3)$ model, ranging from integrable, hence fully regular, to chaotic ones, we have shown that the energy-smoothed wiggliness of the long-time OTOC dynamics and its suppression rate $\alpha$ are robust measures of chaoticity, giving qualitatively the same results for an almost arbitrary choice of OTOC operators.
Due to quantum fluctuations and a rich interplay of regular and chaotic dynamics, smoothing of both the quantum Lyapunov exponent and the wiggliness over neighboring states is necessary for a proper comparison with classical measures of chaoticity. The smoothing reduces the energy resolution; however, it can be improved by increasing the system size, which makes the energy spectrum denser, thus allowing for a narrower smoothing window.

Note that the smoothing procedure mimics, to some extent, taking a superposition of neighboring energy states.
One particular example of such a superposition is a coherent state centered at a given point $q_{j}, p_{j}$.
This state usually has a Gaussian-like local density of states~\cite{Rautenberg2020}. 
Preliminary results show that the wiggliness, in contrast with the quantum Lyapunov exponent, does not depend on whether the initial state is localized in the regular island or chaotic sea in the regime with mixed dynamics.
However, due to the intricate structure of the wiggliness function, its value for the coherent states cannot be directly compared to the wiggliness for the energy eigenstates.
A detailed analysis of the wiggliness for coherent states can be a subject of future work.

The wiggliness is especially relevant in systems of relatively small size, which, due to the shortness of the Ehrenfest time, cannot often reliably capture the exponential rate of the initial OTOC dynamics. In such systems, the wiggliness can effectively supplement the quantum Lyapunov exponent and help distinguish between regular and chaotic quantum dynamics.
The wiggliness can also be used to study quantum chaos in many-body systems without a classical limit; however, further analysis in this field is necessary.

Finally, the short-time evolution requires a reasonable experimental time resolution. 
Therefore, the wiggliness can be experimentally relevant in systems where the initial dynamics is so fast that it cannot be precisely measured. If these systems decohere slowly, the long-time oscillations can be appropriately captured and analyzed. Therefore, they can give valuable information on the system's chaoticity.

\acknowledgments
Pavel Stránský thanks Pavel Cejnar, Lea Santos, Jorge Hirsch, and Jorge Chávez for valuable discussions. Both authors acknowledge the support of the computational server provided by František Knapp.
The work was supported by the Czech Science Foundation (grant no.\,20-09998S) and by Charles University (UNCE/SCI/013).

%


\end{document}